\documentclass[conference]{IEEEtran}
\IEEEoverridecommandlockouts
\usepackage[dvips]{graphicx}
\usepackage{amssymb}
\usepackage{graphicx}
\usepackage[cmex10]{amsmath}
\usepackage{algorithm}
\usepackage{algorithmic}

\usepackage{array}
\usepackage{eqparbox}
\usepackage[tight,footnotesize]{subfigure}
\usepackage{float}
\usepackage{url}
\usepackage{paralist}
\usepackage{multirow}
\usepackage{wrapfig}
\usepackage{color}
\usepackage{tikz}
\usetikzlibrary{shapes.geometric, arrows}
\usepackage{cite}
\usepackage{atbegshi}
\usepackage{setspace}
\usepackage{amsthm}

\newcommand{\RN}[1]{%
	\textup{\uppercase\expandafter{\romannumeral#1}}%
}
\def\BibTeX{{\rm B\kern-.05em{\sc i\kern-.025em b}\kern-.08em
    T\kern-.1667em\lower.7ex\hbox{E}\kern-.125emX}}
\begin{document}

\title{Human-Machine Collaboration for Smart Decision Making: Current Trends and Future Opportunities
}
\author{\IEEEauthorblockN{Baocheng Geng}
\IEEEauthorblockA{\textit{Department of Computer Science} \\
\textit{University of Alabama at Birmingham}\\
Birmingham, USA \\
bgeng@uab.edu}
\and
\IEEEauthorblockN{Pramod K. Varshney}
\IEEEauthorblockA{\textit{Department of Electrical Engineering and Computer Science} \\
\textit{Syracuse University}\\
Syracuse, USA \\
varshney@syr.edu}
}

\maketitle

\begin{abstract}
Recently, modeling of decision making and control systems that include heterogeneous smart sensing devices (machines) as well as human agents as participants is becoming an important research area due to the wide variety of applications including autonomous driving, smart manufacturing, internet of things, national security, and healthcare. To accomplish complex missions under uncertainty, it is imperative that we build novel human machine collaboration structures to integrate the cognitive strengths of humans with computational capabilities of machines in an intelligent manner. In this paper, we present an overview of the existing works on human decision making and human machine collaboration within the scope of signal processing and information fusion. We review several application areas and research domains relevant to human machine collaborative decision making. We also discuss current challenges and future directions in this problem domain.
\end{abstract}

\section{Introduction}

In every aspect of industrial domains and our daily life, ongoing technological advances have led to the large-scale deployments of machines and smart devices empowered with numerous sensing and artificial intelligence (AI)
capabilities. Many tasks and operations previously conducted by humans, e.g., observation acquisition, raw data processing and numerical computation, are now performed by machines. However, since machines are often designed for specific purposes and trained with limited data, they lack the generality and flexibility to be adapted to new scenarios. In critical high-stake situations where human lives and assets are at risk, incorporating human cognitive strengths and expertise in decision making is imperative to improve decision quality and to enhance situational awareness.

Collaborative decision making when the
participants of the decision making process are only machines or electronic devices, e.g., sensors, has been explored extensively
in the signal processing literature in centralized as well
as distributed settings, e.g., \cite{varshney2012distributed,veeravalli2012distributed,kailkhura2015distributed,kay1998fundamentals}. While recent theoretical advances have led to great progress in improving the performance of sensor networks, distributed detection and information fusion, one key challenge of decision making systems is the interaction with the end users, i.e., how to bring humans in the loop for decision making and control. It has become imperative to develop mission-oriented networks consisting of both humans and sensors, i.e., human-machine teams, in semi-autonomous systems where they collaborate with each other to accomplish complex  tasks under uncertainty. According to \cite{hoffman2002rose}, humans surpass machines in their ability to improvise and use flexible procedures, exercise judgement and reason inductively. On the other hand, machines outperform humans in responding quickly, performing repetitive and routine tasks, and reason deductively (including computational
ability). The emerging paradigm of human-machine inference
teams aims to combine complementary cognitive strengths of humans and computation abilities of machines in an intelligent
manner to achieve
higher performance than either humans or machines by
themselves.

\begin{figure}[htb]
	\centering
	\includegraphics[width=1\columnwidth]{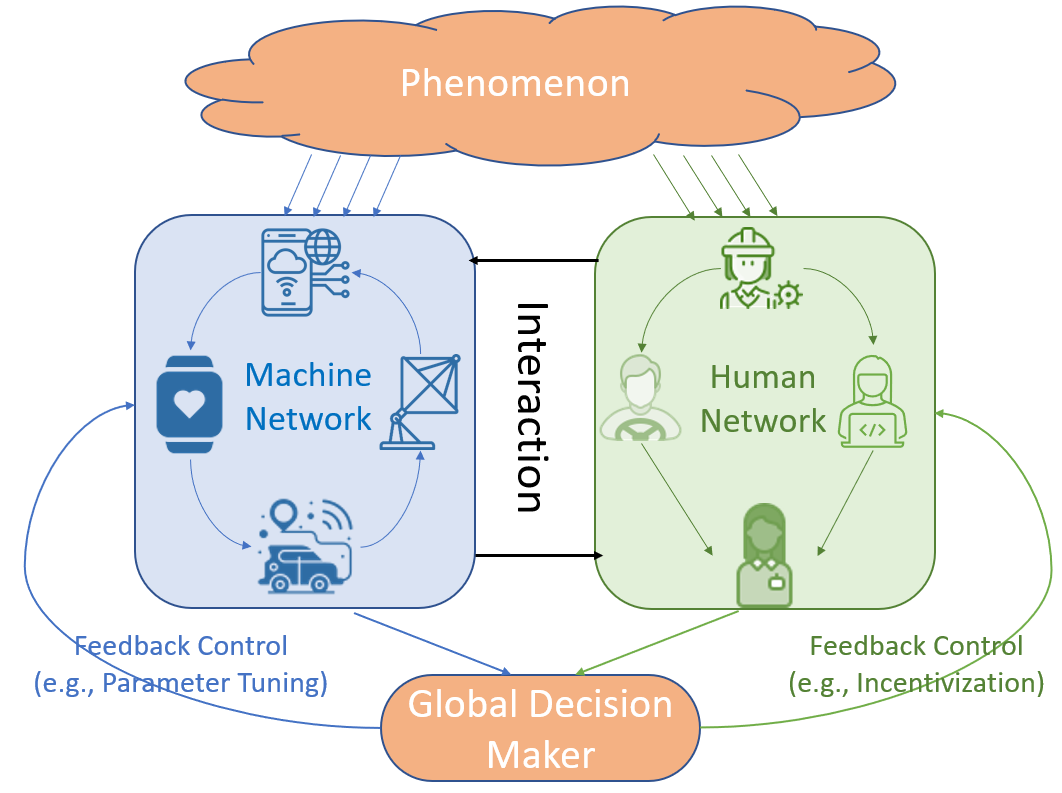}
	\caption{Notional human machine collaboration architecture.}
\end{figure}

In Fig. 1, we show a typical smart decision making architecture where the networks of humans and machines observe the same phenomenon of interest (PoI) from multiple perspectives. In such an inference network, humans can take on a number of different roles: for instance, they can assume the role of {global decision makers}, making judgments based on the outcome of  machine measurements. Alternatively, humans can also operate as `sensors' themselves, e.g, scouts to gather intelligence, where they make {local} decisions that are subsequently relayed to a global decision maker. For the humans to better interact with the machines, implementation of the system needs to consider two aspects: a) the machine outputs affect the behavior, actions and decisions of the humans, and b) the human factors and behavioral properties influence the deployments and  optimal algorithm design of the machines \cite{geng2019decision,geng2021augmented}. 

There has been extensive work on modeling of human decision making both in individual and group formats from the cognitive psychology point of view \cite{sorkin1998group,sorkin2001signal,swets1961decision,glanzer2009likelihood}. In contrast, the focus of this paper is to summarize the research works on human decision making and human machine collaboration from a signal processing and information fusion perspective. We are particularly interested in the integration of statistical signal processing and cognitive psychology theory for inference and smart decision making, especially in the context of distributed detection and multi-agent frameworks.

The rest of the paper is organized as follows. In Section II, we summarize the research works that study human decision making by combining statistical signal detection techniques with cognitive psychology concepts. In Section III, we review the current development of human in the loop decision making frameworks and configurations in different application areas. We discuss the  challenges and future research directions in Section IV and conclude this review paper in Section V. 

\section{Incorporation of Human Factors in Statistical Signal Detection}
One of the biggest challenges faced in human-machine collaboration concerns the formal characterization of human decision making behavior across different tasks and applications. As recently shown by \cite{plonsky2019predicting}, this characterization has much to gain from the integration of theoretical principles found in the cognitive psychology literature. One promising approach, which is the focus in this paper, involves the integration of two decision making modeling traditions that have a long track record of success, namely  \emph{Statistical Signal Detection Theory} \cite{green1966signal,kellen2018elementary,varshney2012distributed}, and  concepts from \emph{ Cognitive Psychology and Behavioral Biases} \cite{kahneman2013prospect,acciarini2020cognitive,cognitivebiasesheet}. In this section, we summarize several research works that study human decision making by incorporating human related factors in the statistical signal detection framework.


\subsection{Random threshold based decision scheme for humans}

Psychology
experiments suggest that individuals often use threshold based
schemes to make local decisions based on what they observe \cite{rhim2012quantization,sorkin1998group,sorkin2001signal,swets1961decision}. In a threshold based decision making framework, the quality of the local decisions depends on the value of
thresholds used by individual agents.
Most of the works in the existing
literature have assumed that the threshold used by individual
agents is a fixed known value. In multi-person decision
making scenarios, however, it is reasonable to assume that the threshold at each agent is a random variable in contrast to assuming a fixed value since different people may have different threshold levels for their sensory signals depending on individual characteristics. 

There have been research
efforts, which attempt to incorporate the random decision thresholds model  into the theory of signal detection to characterize the performance of group decision making.  For example, the authors of \cite{wimalajeewa2013collaborative} analyzed the decision fusion performance when the individual human agents use different thresholds to make local decisions regarding a given PoI while modeling thresholds as random variables. When the number of agents is large, the authors in \cite{wimalajeewa2014asymptotic} derived the performance of the fusion system that fuses individual decisions based on majority rule which does not require the knowledge of individual thresholds. The authors in  \cite{9443353} proposed a hybrid system that consists of multiple human sub-populations, with the thresholds of each sub-population characterized by non-identically distributed random variables, and a limited number of machines (physical sensors) whose exact values of thresholds are known. For such a hybrid system, they derived the asymptotic performance at the fusion center in terms of Chernoff information.  Furthermore, the authors in \cite{wimalajeewa2018integrating} analyzed the conditions under which human
participation can enhance the overall detection performance and designed strategies for physical sensors taking human related factors into account.  They evaluated the fusion performance when the human sensors possess some auxiliary information regarding the PoI that cannot be measured by the machines.

\subsection{Prospect theoretic human decision making}
Extensive work has been done on the modeling of human behavior in decision making under cognitive biases. One of the most systematic
analysis of judgement under biases is the Nobel Prize winning prospect theory (PT) by Tversky and Kahnemann \cite{kahneman2013prospect}. Building upon the solid theoretical
foundation of expected utility theory (EUT), PT has been adopted to model human decision making under uncertainties in a wide
range of application scenarios \cite{barberis2013thirty}.

\begin{figure}[htb]
	\centering
	\includegraphics[width=0.7\columnwidth]{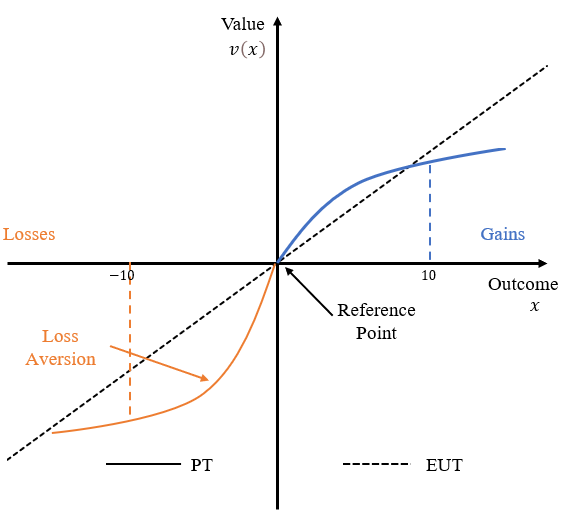}
	\caption{Value function.}\label{fig:ptv}
\end{figure}

According to PT, quantitative outcomes $x$ are represented through the lens of a monotonically increasing \emph{value function} $v(x)$, illustrated in Fig. \ref{fig:ptv}. This value function is convex below a reference point $x_r$, for which $v(x_r) = 0$, and concave for outcomes above it. Following \cite{tversky1992advances}, the mathematical description of the value function is denoted by
\begin{equation}
v(x) = \begin{cases}
(x-x_r)^{\alpha}, & \text{for } x \geq x_r, \\
-\lambda (x_r-x)^{\beta}, & \text{for } x <    x_r,
\end{cases}
 \end{equation}
where $\alpha$ and $\beta$ are risk attitudes coefficients and $\lambda$ is the loss aversion parameter.

\begin{figure}[htb]
	\centering
	\includegraphics[width=0.7\columnwidth]{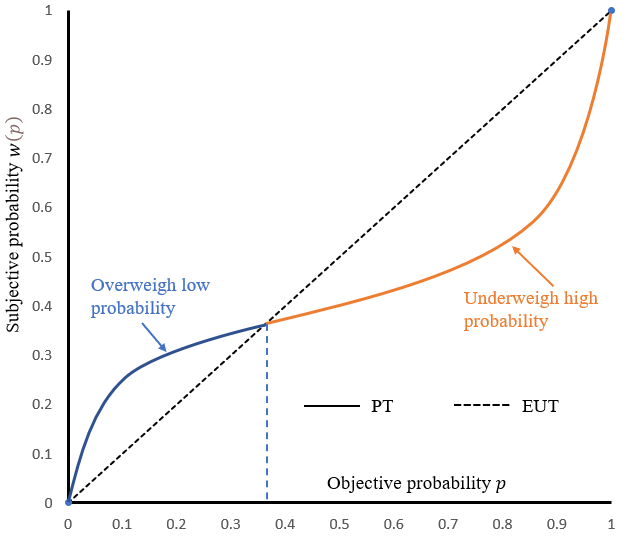}
	\caption{Weighting function.}\label{fig:ptw}
\end{figure}

In turn, subjective probabilities are represented by an inverse-S-shaped weighting function $w(p)$, depicted in Figure \ref{fig:ptw}. This weighting function generally \emph{overweighs} small probabilities and \emph{underweighs} large probabilities. The weighting function is given by 
\begin{equation}
w(p) = \frac{p^\gamma}{(p^\gamma + (1-p)^\gamma)^{\frac{1}{\gamma}}},
\end{equation}
where $w(p)$ gives the subjective probability distorted by the probability distortion coefficient $\gamma$.

Since humans often make decisions in the framework of hypothesis testing and the decision is made by selecting the hypothesis that best supports the given set of observations \cite{poletiek2013hypothesis}, there have been a few works that incorporate PT into hypothesis testing to model human decision making \cite{nadendla2016towards,gezici2018optimality,8976222,8969431}. The optimality of the likelihood ratio test (LRT), which is known to be the optimal decision rule that minimizes the Bayesian risk, was investigated in PT based hypothesis testing in \cite{gezici2018optimality}. The authors showed that the LRT may or may not be optimal for behavioral decision makers under the Neyman-Pearson criterion. In \cite{8976222},  the authors proceeded with utility based approaches, i.e., people make decisions by selecting the choice with larger expected utilities, where they obtained analytical solutions for the optimal decision rule that maximizes the subjective utility of humans under PT. They also considered the optimal configuration design in multi-agent systems where the decision making humans, modeled using PT, are a part of networked human-machine teams.
Moreover, as it has been shown that under PT the decision rules of cognitively biased humans deviate from being optimal (rational), the authors in \cite{8969431} investigated the strategies to ameliorate the biases and help humans achieve better performance while decision making. In particular, three approaches were proposed: i) \textit{modify} the observation presented to a human; ii) \textit{adapt} a physical sensor that helps a human in making the final decision and iii) \textit{adjust} human uncertainties while decision making.

\subsection{Cognitive biases and limitations in human decision making}
In addition to prospect theory, researchers have identified  a  plethora  of (more than 100)  cognitive biases and properties that influence humans' information processing, knowledge
acquisition and decision making \cite{cognitivebiasesheet}. These factors have been incorporated into the statistical signal detection framework to model and analyze human decision making in different contexts. For instance, incorporating cognitive limitations
of humans in group decision making problems was considered
in \cite{rhim2012quantization,varshney2016decision} where categorization schemes based on quantized
prior probabilities for collaborative decision making in a
Bayesian framework were developed. A Bayesian hierarchical  structure was developed in \cite{vempaty2018experiments} to analyze  the decision fusion behavior of humans  at  individual  level,  crowd level and population level. Considering that humans are influenced by the starting point observations or initial beliefs, e.g., anchoring
biases and confirmation bias, the authors in \cite {mourad2016real} studied the problem of
selection, ordering and presentation of data to a human to solve detection problems
As the cognitive working memory of humans is a fairly capacity-limited resource \cite{ecker2014working}, the authors in \cite{9413745} studied how humans make decisions based on internal and external sources of information under cognitive memory limitations. They showed that the order in which information is processed and belief is updated heavily impacts the final decision quality.  Moreover, unlike perfect rational decision maker who always chooses the best action that has the maximum utility,  bounded rational humans will select all the actions in the action space but {\it better options are selected more often}. The authors in \cite{9747866} employed
the bounded rationality model to quantify human uncertainty and evaluated the individual decision making performance when humans have different degrees of bounded rationality.

\section{Relevant Applications and research areas}
A major motivation for the development of human decision making and human machine collaboration is the application areas. These application areas have urgent demands that drive novel architecture design and algorithm solutions on this research topic. We present three critical and timely research areas and applications: crowdsourcing, incentive mechanism design and resource allocation. We discuss
their associated research problems in the context of human machine collaboration.

\subsection{Crowdsourcing}
Crowdsourcing has become an efficient paradigm to utilize distributed human wisdom to collect data annotation for many machine learning applications \cite{burrows2013paraphrase}. There has been growing interest in exploring the influence of crowd workers’ cognitive biases and uncertainties on the quality of their annotations. It is documented that loss aversion affects the crowd workers' effort level and the quality of data annotation \cite{wang2019research,brown2014crowdsourcing}. Other works  show that relevance judgments can be affected by displaying other crowd workers’ judgments (i.e., the bandwagon effect, this is also referred to as Group Think or Herd Behavior) \cite{eickhoff2018cognitive,harris2019detecting}. In different contexts,  several techniques and methods have been proposed to evaluate and mitigate cognitive biases \cite{sabou2014corpus,yue2014weighted,6891807,9133140,8645073,geng2021collaborative,geng2018decision}.  For instance,  coding and decoding algorithms, were developed in order to deal with the uncertainties of the crowd and increase  the classification accuracy \cite{yue2014weighted,6891807}. Based on prospect theoretic analysis of the crowd workers' behavior, the authors in \cite{9133140} proposed a reject option scheme to optimize the performance of classification in the presence of spammers. In the current research context, modeling of human decision making has an important role in the design of efficient and reliable crowsourcing system as it affects each of the following four core concerns: a) task assignment, b) user interface design, c) worker selection and d) privacy protection.

\subsection{Incentive Mechanism Design}
Many of today’s sensing applications
allow users carrying devices with built-in sensors, such as
sensors built in smart phones, and automobiles, to contribute
towards an inference task with their sensing measurements. Since their own resources such as time, energy and processing power will be consumed, the selfish users, e.g., social sensors, may require incentives to participate in the sensing task. A plethora of literature has been dedicated to study game-theoretic solutions to analyze the strategies, behaviors, and interactions of the buyers and sellers in a market based framework. For example,  in \cite{dias2006market}, techniques based on market-theory for task allocation in multi-robot environments are reviewed. In \cite{masazade2013market}, the authors used the
concept of Walrasian equilibrium \cite{walras1954elements} to model market-based
sensor management and dynamic bit allocation problems in energy constrained wireless sensor networks. Moreover, auction based mechanism design has been studied extensively as it serves as an effective framework to manage rules of interaction in the presence of selfish agents within a competitive environment. In particular, researchers have studied the usage of single-sided, two-sided, and combinatorial auctions in different application scenarios such as the internet of things (IoT) \cite{geng2021utility}, target tracking \cite{li2018truthful,9049010,cao2019optimal}, spectrum sensing \cite{jin2018privacy,huang2014truthful}, and cloud computing \cite{zhang2018truthful,sharghivand2021comprehensive}. To design incentive mechanisms that include humans as participants, there are still multiple challenges that need to be addressed such as dynamic change of the market, communication constrained and energy aware mechanisms, privacy and security concerns. 

\subsection{Resource Consumption and Allocation}
In classical resource constrained optimization problems, typically all available resources are consumed so that some system performance metrics are maximized \cite{quan2021strategic,quan2020novel}. However, in most inference tasks, the rate of detection accuracy increment slows down as more resource is spent, i.e, the phenomenon of `diminishing returns' sets in. After a saturation point, the extra profit obtained by spending  additional units of resource is lower than the cost incurred. In this case, the usage of additional amount of resource is neither useful nor advisable. There have been a few works that study humans' strategic decisions for resource consumption and allocation when they are subjected to cognitive biases. For instance, using prospect theory to model humans' decision making behavior under risks, researchers have analyzed and investigated human participants' strategic decisions on information transmission in wireless sensor networks \cite{nadendla2018effects,6747282}, energy exchange and pricing among microgrids \cite{gan2022application,6895275}, optimal energy consumption in binary hypothesis testing, and revenue maximization for the service provider in network slicing \cite{8948257}. In high stake scenarios such as pandemic prevention, natural disaster monitoring, security surveillance and healthcare,  humans make the final decision on resource consumption and allocation, which may consequently affect every aspect in the society.  It is desirable to construct a tool to describe, access, and quantify the impact of humans' cognitive biases on their resource usage strategies, and to mitigate cognitive biases by developing human machine collaboration teams that provide personalized suggestion. 

\section{Challenges in human machine collaboration }

The perceptions, decisions and actions of humans are complex and uncertain processes that represent the intricate interplay between the psychological activity within humans and the influence of outside environment. {A fully cognitive human-machine  collaborative decision making architecture requires the machine to have the ability to understand, anticipate,
and augment the performance of the human; and the human to be able to support, supervise, and improve the operations conducted by the machine  \cite{grigsby2018artificial}. To successfully design a truly interactive symbiosis where humans and machines are tightly coupled together in smart decision making, several  open problems include, but not limited to the following.}

\begin{enumerate}
    \item {}\textbf{Behavioral informatics:}  A better understanding of human behavior and behavioral informatics is fundamental to the design of human machine collaboration teams for smart decision making. 
    It is necessary to dive deeper into psychology and characterize how human behavior is impacted by time constraints, memory limitations, emotion state as well as stimulus from the outside environment. To achieve `order of magnitude increases in available, net thinking power resulting from
linked human-machine dyads' \cite{schmorrow2004augmented}, it becomes imperative to perform human cognitive state sensing for designing efficient communication interfaces between the human and the machine. Relevant topics includes the real time prediction of human cognitive workload based on sensor-based brain signals as well as the design of system augmentations such as  offloading tasks or  assisting users with modality-specific support \cite{wang2021taking,afergan2014dynamic}.

\item{}\textbf{Trust among humans and machines:} In human-machine teams, the authors in \cite{mayer1995integrative} have defined trust as `the willingness of a party to be vulnerable to
the actions of another party based on the expectation that the other will perform a
particular action important to the trustor, irrespective of the ability to monitor or control
that other party'.  In particular, many machine learning and control systems developed in a wide variety of applications are black boxes that do not explain how the decisions are made. The challenge is to develop a quantitative definition of trust and establish clear guidelines to construct human-machine transparency and enhance calibrated trust between the human and the machine.

\item{}\textbf{Situational awareness:} Situational awareness (SA) refers to the user's familiarity of the task environment, the perception of the task status, and the anticipation of future states. If humans are not appropriately included in the loop, it is very likely that the human is not aware of or not familiar with the machine's  task operation.  In such a situation where there is over-reliance on machine automation, the human's understanding of the work environment, i.e., SA, is jeopardized. The loss of SA (also referred to as complacency or automation induced decision biases in different works) compromises the human's level of expertise and ability to perform the
automated tasks manually in case of unpredictable automation failure and it may cause severe breakdown in critical applications like autopilot and submarine navigation systems. Hence, the concerns of SA must be addressed in the design of human-machine collaboration to prevent irreparable disasters.

\item{}\textbf{Herding, nudging and incentives:} Humans are also known to be subject to herding and nudging phenomena. To elicit desirable outputs from humans, future research work can  proceed with some explorations along these lines. a) The optimum design and task allocation of collaborative human-machine networks \cite{sabou2014corpus,ipeirotis2011managing}. This will include change in strategies of individual nodes, e.g., adapting the threshold of some or all the nodes or shaping the input to selected nodes during the inference process. b) The suitable distribution of the tasks and workload to be performed by humans and machines leading to semi-autonomous systems \cite{sriranga2020human}. c) Another important consideration will be the incentivization  measures of humans to actively engage in the inference process, which can be posed in a reinforcement learning based framework. 
\end{enumerate}

\section{Conclusion}

In almost all decision making systems developed in signal processing and machine learning, the objective function is based solely on objective performance measures and is devoid of any human perception considerations. The incorporation of human in the loop will generalize cognitive systems by allowing humans and machines being tightly coupled in the same working environment for advanced interaction. 

In this paper, we reviewed several research works that combine statistical signal processing and cognitive psychology theory to solve inference problems in different contexts.
Development of the next generation systems for collaborative human-machine decision  making in complex environments requires inputs from different disciplines such as  statistical signal processing, artificial intelligence (AI), machine learning (ML), economics, experimental psychology, and neuroscience. The ultimate goal is to merge the best of humans with the best of machines so that humans and machines can interact and complement each other while engaging in different decision making and AI tasks.
\bibliography{refer}
\bibliographystyle{IEEEtran}
\end{document}